\documentclass[a4paper, amsfonts, amssymb, amsmath, preprint, nofootinbib, twoside,pra]{revtex4}
\usepackage[english]{babel}
\usepackage[utf8]{inputenc}
\usepackage[colorinlistoftodos, color=green!40, prependcaption]{todonotes}
\usepackage{amsthm}
\usepackage{mathtools}
\usepackage{physics}
\usepackage{xcolor}
\usepackage{graphicx}
\usepackage{adjustbox}
\usepackage{placeins}
\usepackage[T1]{fontenc}
\usepackage{lipsum}
\usepackage{csquotes}
\usepackage[hidelinks]{hyperref}
\begin{document}
\title{Analytical approach for pure high, even-order dispersion solitons}

\author{Xing Liao}

\author{Jiahan Huang}

\author{Daquan Lu}
\email{ludq@scnu.edu.cn}
\author{Wei Hu}
\email{huwei@scnu.edu.cn}

\affiliation{Guangdong Provincial Key Laboratory of Nanophotonic Functional Materials and Devices, \\ School of Information and Optoelectronic Science and Engineering, \\ South China Normal University, Guangzhou 510631, China}

\begin{abstract}
We theoretically solve the nonlinear Schr\"{o}dinger equation describing the propagation of pure high, even order dispersion (PHEODs) solitons by variational approach. The Lagrangian for nonlinear pulse transmission systems with each dispersion order are given and the analytical solutions of PHEOD soltions are obtained and compared with the numerical results. It is shown that the variational results approximate very well for lower orders of dispersion ($\le 8$) and get worst as the order increasing. In addition, using the linear stability analysis, we demonstrate that all PHEOD solitons are stable and obtain the soliton internal modes that accompany soliton transmission. These results are helpful for the application of PHEOD solitons in high energy lasers.
\end{abstract}

\maketitle

\section{Introduction}
Conventional temporal solitons are wave packets generated by the balance of nonlinear Kerr effects and group velocity dispersion~\cite{Hasegawa1973APL,Agrawal2019book}.
For those solitons, higher orders of dispersion are considered to be a perturbation that limits transmission capacity, causing energy loss and increasing soliton instability~\cite{Akhmediev1994OC,Karlsson1994OC}, which is regarded as a nuisance that needs to be suppressed or managed.
However, the experimental discovery of pure quartic solitons (PQSs)~\cite{Blanco2016NC}, which are generated purely from the balance of the anomalous quartic dispersion and nonlinear effect, reveals that higher order dispersion can be a beneficial factor for the formation of some solitons.
More recently, the concept of PQSs has been further extended to higher orders~\cite{Runge2021PRR}.
Each even-order dispersions can be balanced with nonlinear effects, realizing pure even-order dispersion (PHEOD) solitons.
Unlike conventional solitons, PHEOD solitons are with approximate Gaussian shape at center and decaying oscillatory tails at edges~\cite{Tam2019OL,Runge2021PRR}.
At the same bandwidth, the PHEOD solitons have higher energy and a flatter spectral distribution than conventional solitons~\cite{Taheri2019OL,Runge2021PRR,Blanco2021APL}.
These properties make them promising for practical applications in high power soliton lasers~\cite{Runge2020NP,Zhang2024NC} and frequency combs~\cite{Taheri2019OL}.
Besides, the unique properties of PHEOD solitons have also sparked further research on related solitons with mixed high orders of dispersion~\cite{Qiang2022PRA,Qiang2022JPA}.

The previous study of PHEOD solitons has mainly focused on experiments and numerical simulations~\cite{Runge2021PRR}, but there is a lack of analytical studies on the fundamental theoretical aspects of the PHEOD solitons.
The variational approach~\cite{Manton1979NPB,Malomed2002book} is undoubtedly one of the most useful methods for constructing approximate analytical solutions to the nonlinear Schr\"{o}dinger equation.
It was first introduced by Anderson for the analysis of nonlinear pulse propagation in optical fibers~\cite{Anderson1983PRA}.
Subsequently, Karlsson et al. investigated the propagation characteristics of pulses in a nonlinear optical fiber containing negative fourth-order dispersion and anomalous second-order dispersion using the variational approach~\cite{Karlsson1994OC}.
In 2019, Taheri et al. demonstrated the existence of dissipative PQSs sequences in Kerr microresonators dominated by quartic dispersion using the Lagrangian variational approach~\cite{Taheri2019OL}.
More recently, the analytical solution of PSQ~\cite{LiYH2022OC} and the effective interaction potential between PQSs~\cite{DaiJX2022CSF} was studied based on the variational approach.

In this paper, we theoretically study the analytical solution of PHEOD solitons by the variational approach.
Based on the Lagrangian of the system for different dispersion order, we obtain the approximate analytical solutions of PHEOD solitons and compares with the numerical results.
In addition, by calculating the linear-stability eigenspectrum of PHEOD solitons, we demonstrate that all solitons are stable and obtain the soliton internal modes that accompany soliton transmission.

\section{The Variational Approach}

The propagation of the optical pulse with envelope $\Phi(z, t)$ in a fiber with only $m$th-order dispersion can be described by the PHEOD nonlinear Schr\"{o}dinger equation (NLSE), \cite{Runge2021PRR,Kevin2020PRA}
\begin{equation}
	 i\frac{\partial \Phi}{\partial z}+(i)^{m}\frac{\beta_m}{m!}\frac{\partial^{m}\Phi}{\partial t^{m}}+\gamma|\Phi|^2\Phi=0,
	\label{NLSE}
\end{equation}
where $z$ and $t$ are the propagation distance and time coordinates, $\gamma$ and $\beta_m$ are the nonlinear parameter and $m$th-order dispersion, where $m$ is an even integer. By introducing the normalized variables, 
\begin{equation} \label{ct}
    u=\sqrt{\gamma L_m } \Phi, \quad \zeta = \frac{z}{L_m}, \quad \tau = \frac{t}{T_0}.
\end{equation}
Equation~(\ref{NLSE}) can be rewritten as 
\begin{equation}
    i\frac{\partial u}{\partial\zeta}+{\rm{sgn}}(\beta_{m})\frac{i^m}{m!}\frac{\partial^{m} u}{\partial \tau^m}+|u|^2u=0.
    \label{NLSEs}
\end{equation}
Here $T_0$ is the initial pulse width, $L_m=T_{0}^{m}/|\beta_m|$ is defined as the $m$th-order dispersion length. In the following of this paper, we choose ${\rm{sgn}}(\beta_m)=-1$ and focus on the case of anomalous $m$th-order dispersion fiber system, in which PHEOD solitons exist.

It is known that NLSE is equivalent to a variational problem:
\begin{equation}
     \delta \int_{0}^{\infty} \!\!\! \int_{-\infty}^{\infty}\!\!\! \mathcal{L} 
     d\tau d\zeta=0,
    \label{BF}
\end{equation}
where $\mathcal{L}$ is the Lagrangian density. The key point to a variational problem is to find the proper expression of  Lagrangian density.  For PHEOD NLSE, refer to the Lagrangian density in Ref.~\cite{ChenMN2019CNS}, we give:  
\begin{equation}
     \mathcal{L}=\frac{i}{2}\left(u^{\ast}\frac{\partial u}{\partial \zeta} -u\frac{\partial u^{\ast}}{\partial \zeta}\right)-\frac{1}{m!} \left|\frac{\partial^{n}u}{\partial \tau^{n}}\right|^{2} +\frac{1 }{2} |u|^{4},
    \label{LGMD}
\end{equation}
where the asterisk denotes the complex conjugate, $u_\zeta=\partial u/\partial \zeta$, $u^{(n)}_{\tau}=\partial^{n} u/\partial \tau^{n}$, and $n=m/2$ is a integer. For each integer $m$  and $n$,
the Eq.~(\ref{NLSEs}) can be restated from the Euler-Lagrange equation   
\begin{equation}
     \frac{\partial \mathcal{L}}{\partial u^{\ast}} -\frac{\partial}{\partial \zeta} \left(\frac{\partial \mathcal{L}}{\partial u^{\ast}_\zeta} \right) +(-1)^{n}\frac{\partial^{n}} {\partial \tau^{n}} \left( \frac{\partial \mathcal{L}} {\partial u^{\ast(n)}_{\tau}} \right) =0.
    \label{OL}
\end{equation}
Equations (\ref{LGMD}) and (\ref{OL}) are the fundamental of this paper  and valid for all pure even-order dispersion. They may be reduced to the Lagrangian density and  Euler-Lagrange equation for  conventional solitons (with 2nd order group velocvity dispersion) or PQS (with the quartic dispersion). 

To obtain an approximate solution for Eq.~(\ref{NLSEs}) by the variational approach, one have to find a suitable trial function to approximate the  exact solution.  Since the main part of a PHEOD soliton is Gaussian-like, we adopt the Gaussian function as the trial solution \cite{Blanco2021APL,Runge2021PRR}, i.e.
\begin{equation}
    u(\zeta,\tau)= A(\zeta)\exp[ib(\zeta)]\exp\left[-\frac{[1+iC(\zeta)]\tau^{2}}{2w^{2}(\zeta)}\right],
    \label{ST}
\end{equation}
where $A$, $w$, $C$, and $b$ are the amplitude, the pulse duration, the frequency chirp, and the entire phase, respectively. Substituting the ansatz Eq.~(\ref{ST}) into Eq.~(\ref{LGMD}) and integrating it with respect to the time coordinate $\tau$ yields the Lagrangian $L=\int_{-\infty}^{\infty} \mathcal{L}d\tau$, 
\begin{eqnarray}
    L &=&  \sqrt{\pi}A^2 w \left[\frac{\sqrt{2}}{4}A^{2}-\frac{db}{d\zeta} +\frac{1}{4}\frac{dC}{d\zeta} \right. \nonumber\\
    & &- \left.\frac{C}{2w}\frac{d w}{d\zeta}-\frac{(1+C^2)^n}{n!2^{m}w^{m}}\right].
    \label{LGLRL}
\end{eqnarray}

The reduced variational problem $\delta \int_0^{\infty} L d\zeta=0$ corresponds to a set of evolution equations for generalized coordinates $A$, $w$, $C$, $b$, i.e.
\begin{equation}
    \frac{d}{d\zeta}\left(\frac{\partial L }{\partial \dot{q}}\right)- \frac{\partial L }{\partial q}=0,
    \label{New}
\end{equation}
where $\dot{q_j}=d q_{j}/d\zeta$, $q_j=A, w, C, b$, respectively. Then Eq.~(\ref{New}) yields four ordinary differential equations,
\begin{eqnarray}
\frac{C}{2w} \frac{dw}{d\zeta} \!&=&\! \frac{1}{4}\frac{dC}{d\zeta} -\frac{db}{d\zeta} + \frac{\sqrt{2}}{2} A^2 -\frac{(1+C^2)^n}{n! 2^m w^{m}} ,\label{Avps}
  \\
\frac{C}{2w}\frac{d w}{d\zeta} \!&=&\! \frac{3}{4}\frac{d C}{d\zeta}- \frac{db}{d\zeta} +\frac{\sqrt{2} A^2}{4} + \nonumber \\
     & & +\frac{(m-1)(1+C^2)^n}{n! 2^m w^{m}},\label{Wvps}
  \\
    \frac{dw}{d\zeta}\!&=&\! - \frac{nC(1+C^2)^{n-1}}{n! 2^{m-2} w^{m-1}},\label{Cvps}
    \\
    & & \frac{d}{d\zeta}(\sqrt{\pi}w A^{2}) = 0.\label{Bvps}
\end{eqnarray}

Equation (\ref{Bvps}) represents the energy conversation of the pulse, i.e., the  pulse energy $E=\int_{-\infty}^{\infty}|u(\zeta,\tau)|^2d\tau=A^{2}w\sqrt{\pi}$ is unchange during propagation. Therefore Eqs.~(\ref{Avps})-(\ref{Cvps}) can be   simplified as
\begin{eqnarray}
    \frac{db}{d\zeta} \!\!&=&\!\! \frac{5\sqrt{2}E}{8\sqrt{\pi}w} - \frac{(m+2)(1+C^2)^n}{n! 2^{m+1} w^{m}} 
    \nonumber  \\
     & &+\frac{nC^2(1+C^2)^{n-1}}{n! 2^{m-1} w^{m}},\label{Byanhuas}
    \\
    \frac{d C}{d\zeta} \!\!&=&\!\! \frac{E}{\sqrt{2\pi}w}-\frac{m(1+C^2)^n}{n! 2^{m-1} w^{m}},\label{Cyanhuas}
    \\
    \frac{dw}{d\zeta} \!\!&=&\!\! - \frac{nC(1+C^2)^{n-1}}{n! 2^{m-2} w^{m-1}}\label{Wyanhuas}.
\end{eqnarray}
This three equations together describe the evolution of the PHEOD solitons. The evolution equations  are identical as that for conventional solitons \cite{Malomed2002book} or PQS \cite{LiYH2022OC} as $m$ is set to $2$ or $4$, respectively.

\section{Solution of PHEOD solitons}

For a stationary soliton solution, we have $dA/d\zeta=0$ and $C=0$. Therefore, the analytic solution of the PHEOD Soliton can be written as
\begin{equation}
    u(\zeta,\tau)= \sqrt{P_s}\exp\left(-\frac{\tau^{2}}{2w_{0}^{2}}\right) e^{i\mu_s\zeta},
    \label{Analytic}
\end{equation}
where $w_0=w(0)$ is the initial pulse duration, $\mu_s=db/d\zeta$ is the propagation constant, and $P_s=A^2$ is the peak power of the soliton. based on Eqs.~(\ref{Byanhuas})--(\ref{Wyanhuas}), the value of those parameters are given as
\begin{eqnarray}
    \mu_s &=& \frac{2m-1}{n! 2^{m} w_{0}^{m}}\label{mus}, \\
     P_s &=& \frac{m}{\sqrt{2}n! 2^{m-2}w_{0}^{m}}\label{Ps} \\
	 E_s &= & \frac{\sqrt{2\pi}m}{n! 2^{m-1} w_{0}^{m-1}}\label{Es}.
\end{eqnarray}
Here $E_s$  is the  total energy of solitons. 

\begin{figure}[tbh]
		\centering
		\includegraphics[width=10cm]{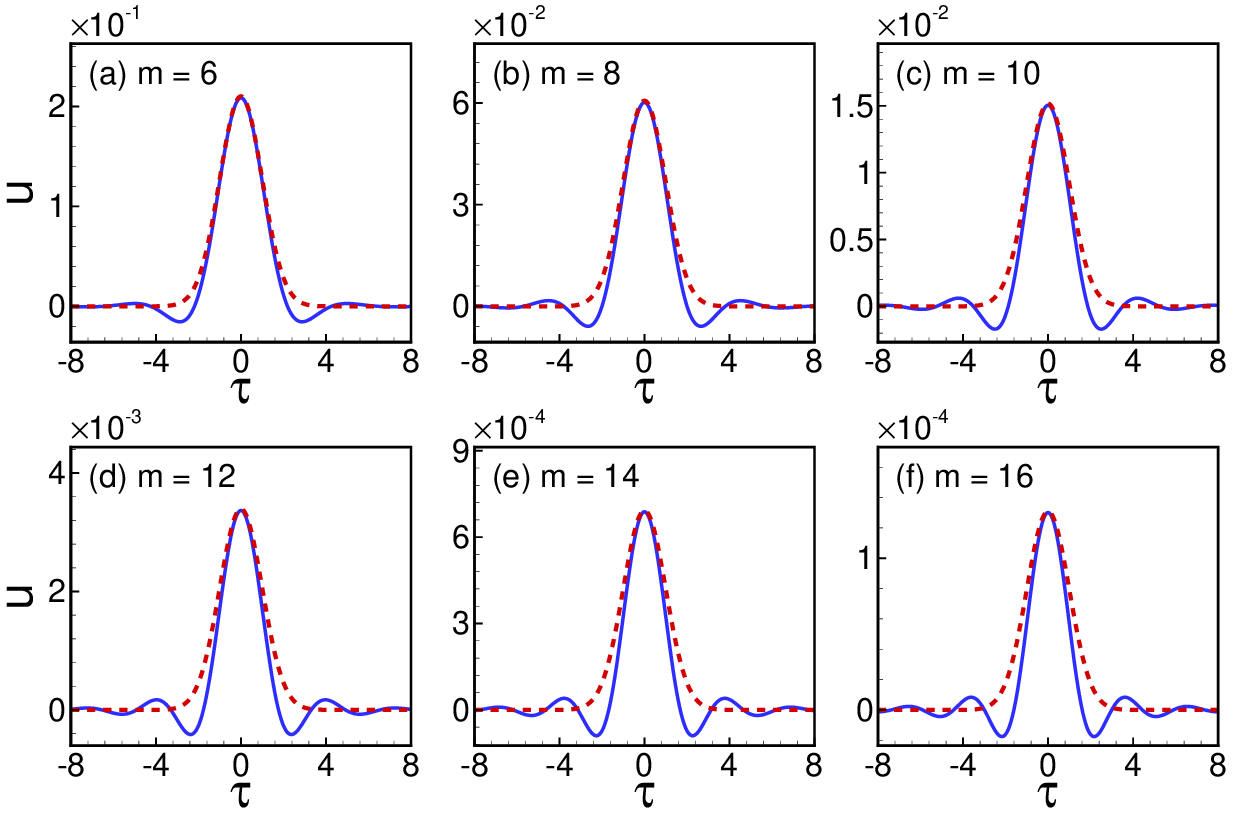}
		\caption{Comparison of the analytic and numerical solutions of the PHEOD solitons for different order with same value of propagation constant $\mu_s$. }\label{fig1}
\end{figure}

To validate the analytic solution Eqs.~(\ref{Analytic})--(\ref{Es}), we compare the results with  the numerical solutions obtained based on Eq.~(\ref{NLSEs}) using the  modified squared-operator
method (MSOM) \cite{Yang2007}. Here the propagation constant $\mu_s$ in the MSOM program is set based on Eq.~(\ref{mus}) with $w_0=1$. The numerical and analytic solutions are shown in Fig.~\ref{fig1} as solid and dashed curves, respectively, for comparison. It is known that the PHEOD solitons have the Gaussian shape at the center and the oscillatory decaying tails at the edge \cite{Runge2021PRR}. One can see that the analytic solutions approximate well with its numerical counterparts in the center part, with much similar amplitude $\sqrt{P_s}$,  and slightly narrow pulse durations. It means the analytic relation between peak power $P_s$ and propagation constant $\mu_s$, i.e.
\begin{equation} \label{PPP}
  P_s=\frac{\sqrt{8}m}{2m-1}\mu_s,
\end{equation}
is valid with high accuracy.

From Fig.~\ref{fig1}, the pulse width of the numerical solutions are narrower than that of the analytic solutions. The difference get more obvious as the dispersion order increasing. Because of the oscillatory tails existing in the numerical soliton sultions, the Gaussian width (at the $1/e$ maximum) is not proper for PHOED solitons. Therefore, the full width at half maximum (FWHM width), which is commonly used in experiment, is suitable here for comparison. 

\begin{figure}[tbh]
		\centering
		\includegraphics[width=10cm]{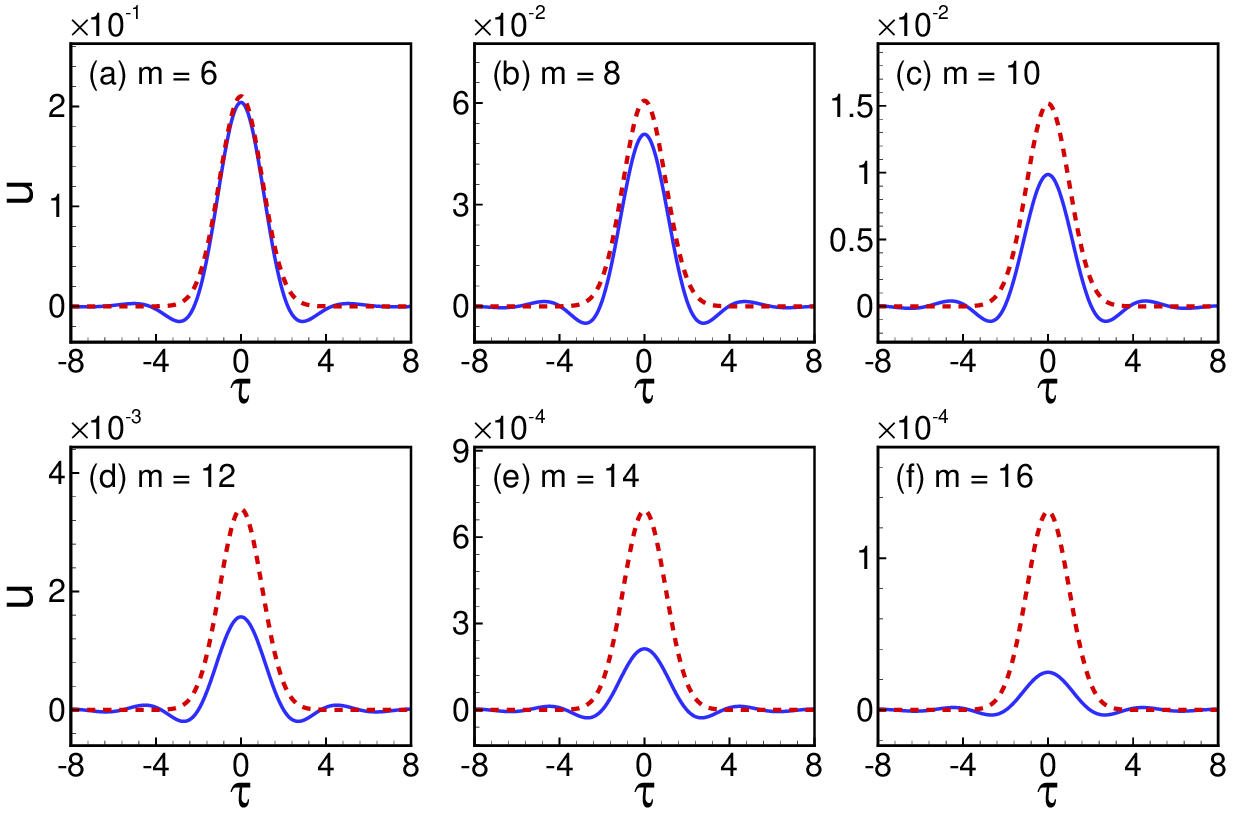}
		\caption{Comparison of the analytic and numerical solutions of the PHEOD solitons for different order with same value of FWHM pulse width.}\label{fig2}
\end{figure}

The numerical and analytic solutions with same FWHM width are compared shown in Fig.~\ref{fig2}.  Here the FWHM width for all solutions have the same value of $1.6651=2(\ln 2)^{1/2}$, corresponding to Gaussian width $w_0=1$. So the analytical results in Fig.~\ref{fig1} and Fig.~\ref{fig2} are identical. The numerical results in Fig.~\ref{fig1} and Fig.~\ref{fig2} have same pulse shape for each dispersion order, and belong to a same continue family of PHOED solitons with different pulse widths and amplitudes, which can be parameterized by the value of $\mu_s$. 
One can see in Fig.~\ref{fig2} that the difference of the peak power increases dramatically as the dispersion order increasing.  This may be caused by the high-order power function in the scale transformation~\cite{Runge2021PRR}, i.e., a slight difference of pulse width in Fig.~\ref{fig1} lead to a large difference of pulse amplitude in Fig.~\ref{fig2}.

Next, for comparing with the experimental data, the analytical results Eqs.~(\ref{Analytic})--(\ref{Es}) in the dimensionless form are converted into the  physical dimension based on Eq.~(\ref{ct}), and the Gaussian width $w_0$ is replaced by the FWHM width $w_F=2(\ln 2)^{1/2}w_0$, yields 
\begin{eqnarray}
	\mu_{0}\!\!&=&\!\!\frac{\gamma P_0}{\sqrt{8}}(2-\frac{1}{m}),\label{Constant} 
    \\
    E_{0}\!\!&=&\!\!\frac{m(\ln 2)^{\frac{m-1}{2}} \sqrt{2\pi}|\beta_m|}{n!\gamma T^{m-1}_0}     \equiv \frac{M_{m}|\beta_m|}{\gamma T^{m-1}_0},\label{Eneargy}
    \\
    P_0\!\!&=&\!\!\frac{2\sqrt{2}m(\ln 2)^{\frac{m}{2}}|\beta_m|}{n! \gamma T^{m}_0} \equiv \frac{N_{m}|\beta_m|}{\gamma T^{m}_0},\label{Peak power}
\end{eqnarray}
where
\begin{eqnarray}
	M_{m}\!\!&=&\!\!\frac{m(\ln 2)^{\frac{m-1}{2}}\sqrt{2\pi}}{n!},\label{M}
	\\
	N_{m}\!\!&=&\!\!\frac{2\sqrt{2}m(\ln 2)^{\frac{m}{2}}}{n!}.\label{N}
\end{eqnarray}
Here we obtain the  analytical expression of the key parameters: the relation between peak power
and nonlinear phase shift ${\mu_{0}}/({\gamma P_0})$, the energy coefficient $M_m$, and the peak power coefficients $N_m$, which have been found numerically in Ref.~\cite{Runge2021PRR}.

\begin{table*}[htb]
	\caption{ Comparison of key PHEOD soliton parameters for each dispersion orders $m$. The numerical results are same as in Table. 1 of Ref.~\cite{Runge2021PRR}}
	\begin{tabular}{ccccccc}\hline \hline
		$m$&\multicolumn{2}{c}{$\mu_{0}/\gamma P_0$}&\multicolumn{2}{c}{$M_m$}&\multicolumn{2}{c}{$N_m$}\\
		& \quad analytical \quad & \quad numerical \quad & \quad analytical \qquad &\qquad numerical\qquad &\qquad analytical\qquad &\qquad numerical\\
\hline
		4 & 0.618 & 0.620 & 2.89 &   2.87 &   2.72 &   2.71\\
		6 & 0.648 & 0.658 & 1.00 &  0.94 &  0.94  & 0.90\\
		8 & 0.662 & 0.674 & 0.231 & 0.162 & 0.217 & 0.156\\		
    10 & 0.671 & 0.682 & $4.01\times10^{-2}$ & $1.78\times10^{-2}$ & $3.77\times10^{-2}$ & $1.64\times10^{-2}$ \\
    12 & 0.677 & 0.687 & $5.56\times10^{-3}$ & $1.16\times10^{-3}$ & $5.22\times10^{-3}$ & $7.02\times10^{-4}$ \\
    14 & 0.681 & 0.690 & $6.42\times10^{-4}$ & $5.86\times10^{-5}$ & $6.03\times10^{-4}$ & $5.78\times10^{-5}$ \\
    16 & 0.685 & 0.693 & $6.36\times10^{-5}$ & $2.50\times10^{-6}$ & $5.98\times10^{-5}$ & $2.17\times10^{-6}$ \\
		\hline\hline
	\end{tabular}
	\label{tabparameter}
\end{table*}

In Tab.~\ref{tabparameter}, we compare our analytical results of three key PHEOD soliton parameters with the corresponding numerical results in Ref.~\cite{Runge2021PRR}. For the relation between peak power and nonlinear phase shift, i.e. ${\mu_{0}}/({\gamma P_0})$, which is same as Eq.~\ref{PPP}, it is shown the analytical results approximate well to but are slightly smaller than the numerical result. 
For the  energy coefficient $M_m$  and the peak power coefficients $N_m$, the analytical results are only sightly larger than the numerical results  at lower orders of dispersion ($m<8$).
But at higher orders of dispersion ($m>8$), the deviation between the two becomes larger and larger with the increase of $m$.

\section{Linear Stability Analysis of PHEOD solitons}

\begin{figure}[htbp]
	\centering
	\includegraphics[width=7cm]{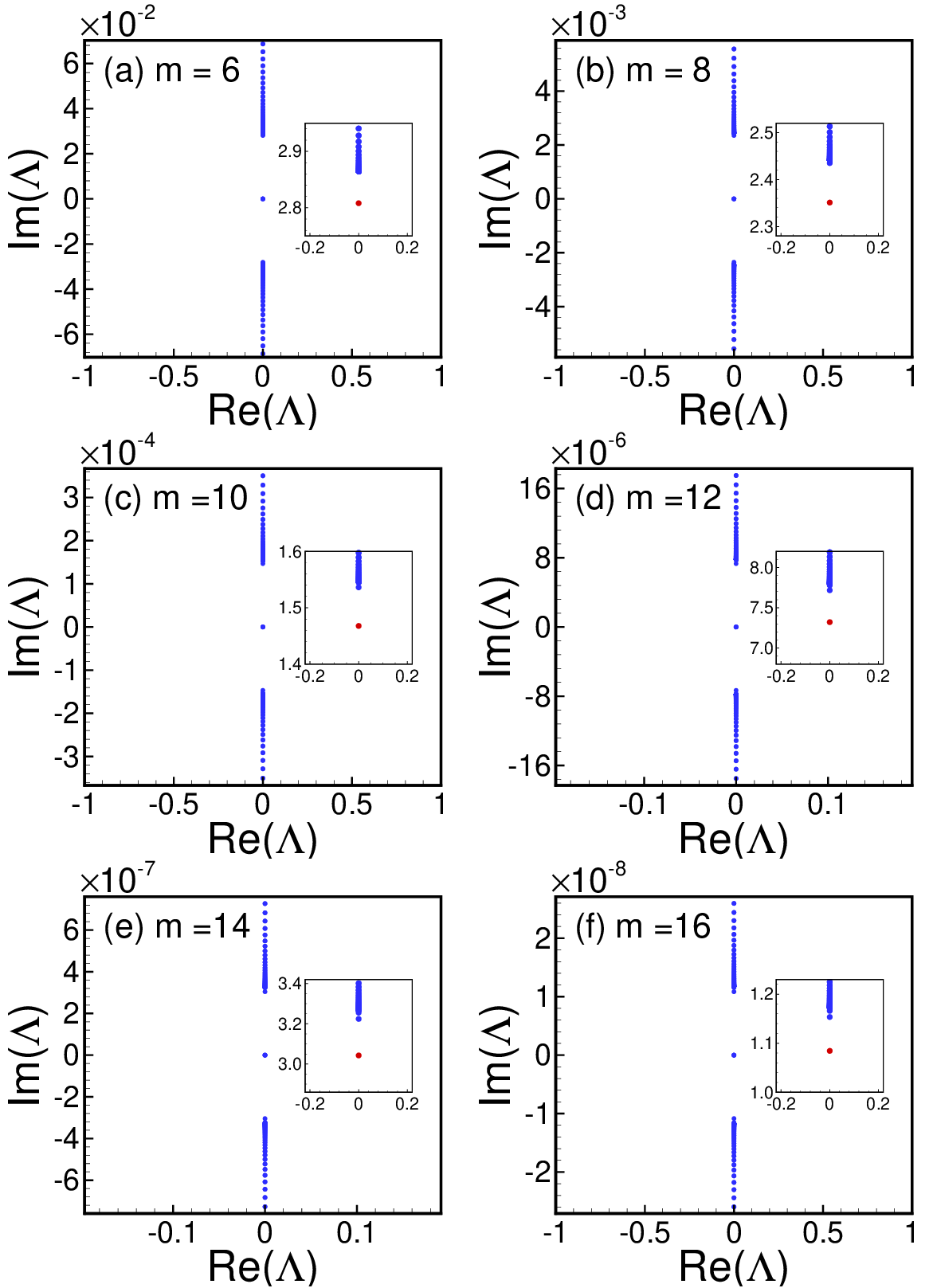}\\
    \caption{Linear stability spectrum of the PHEOD solitons for different dispersion orders. }\label{fig3}
\end{figure}

Now let us further study the stability of the PHEOD soliton. We apply small perturbations to the stationary PHEOD soliton~\cite{Tam2019OL}, i.e.,
\begin{equation}\label{WR}
	u(\tau,\zeta)=\left[u_0(\tau)+f(\tau)e^{\Lambda \zeta}+g^{\ast}(\tau)e^{\Lambda^{\ast} \zeta}\right]e^{i\mu_s \zeta},
\end{equation}
where $u_0(\tau)$ is the stationary PHEOD soliton solution numerically  obtained based on Eq. (3) using the MSOM \cite{Yang2007}, $f$ and $g$ are the  perturbations,
  $\Lambda$ is the eigenvalue of $f$ and $g$.
  Substituting Eq.~(\ref{WR}) into  Eq.~(\ref{NLSEs}) and ignoring the nonlinear terms yields the eigenvalue equations:
\begin{eqnarray}
	\left[-\frac{(i)^{m}}{m!}\frac{\partial^{m} }{\partial \zeta^{m}}-\mu_s+2 u^2_0 \right] f+ u^2_0g \!\!&=&\!\! -i\Lambda f,\label{s1}
	\\
	- u^2_0 f-\left[-\frac{(i)^{m}}{m!}\frac{\partial^{m} }{\partial \zeta^{m}}-\mu_s-2 u^2_0 \right] g \!\!&=&\!\! -i\Lambda g.\label{s2}
\end{eqnarray}
By using the Fourier collocation method, the eigenvalues  for Eqs.~(\ref{s1}) and (\ref{s2}) with dispersion order up to 16 were gotten numerically, as shown in Fig.~\ref{fig3}. It can be seen that all eigenvalues are imaginary, which indicates that  no  perturbation grow exponentially and  thereby all the PHEOD solitons are stable. 

\begin{figure}[htbp]
		\centering
		\includegraphics[width=7cm]{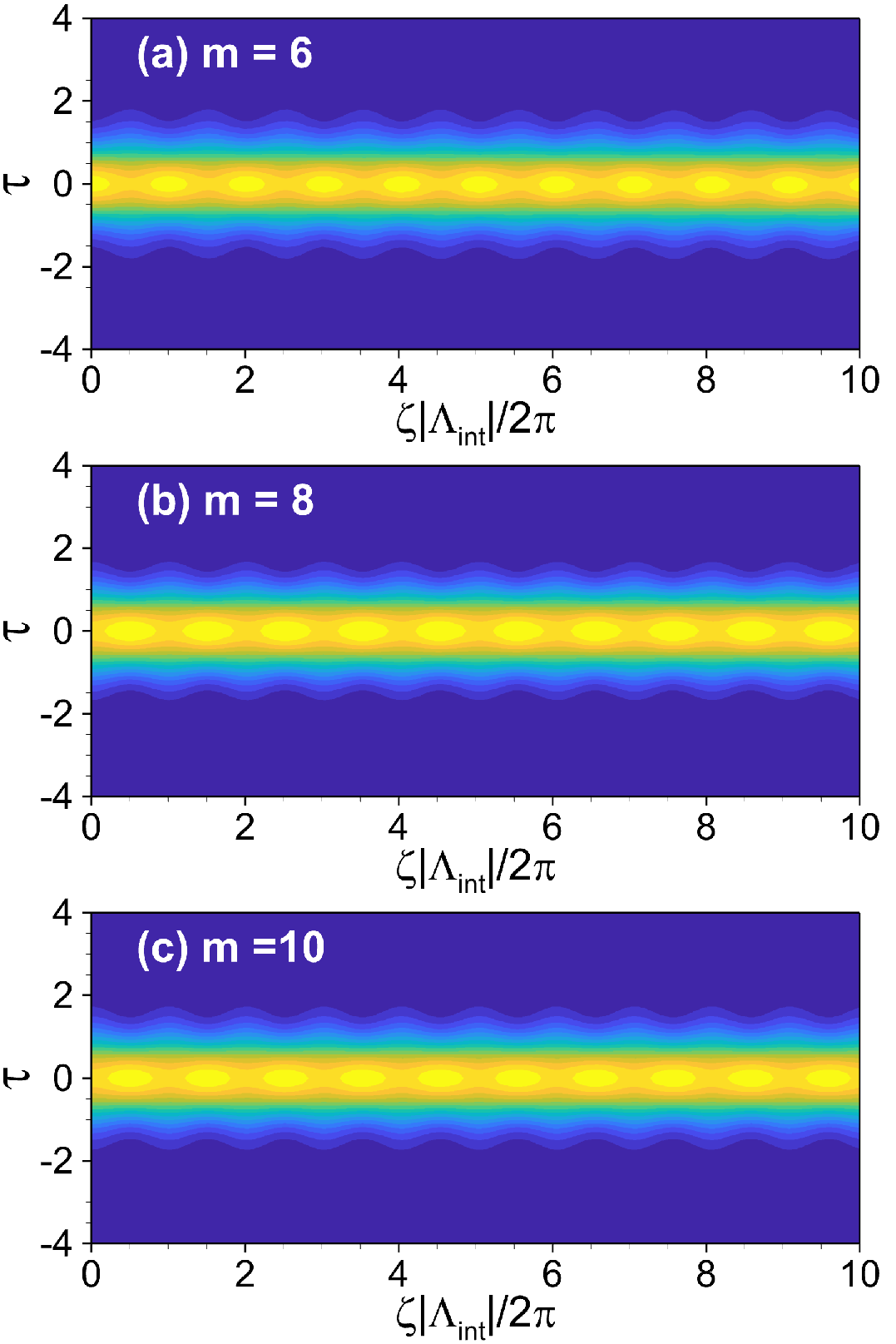}\\
		\caption{Evolutions of PHEOD solitons subjected to a  10\% perturbation by their internal modes, for $m = 6, 8, 10$.}
\label{fig4}
\end{figure}

The linear stability spectrum for each PHEOD soliton always appear in complex conjugate pairs , and contains a zero eigenvalue of multiplicity four, that reflect the phase invariance of the PHEOD solitons.  The inset figures in  Fig. \ref{fig3} show that there is a pair of pure imagery eigenvalues $\pm i\Lambda_ {int}$, and rest of the spectrum is continuous eigenvalues. The edge of continuous spectrum is at $\pm i \mu_{s}$. The eigenvalues $\pm i \Lambda_{int}$ present the internal mode, which can co-propagate with the solitons and results in breathing of soliotns.  Fig.~\ref{fig4} shows the evolution of PHEOD solitons subjected to a  10\% perturbation by their internal modes,  undergoing the expected period of $2\pi/|\Lambda_ {int}|$. The evolutions of the breathing solitons with internal mode are all similar for PHEOD solitons with diepserion order up to $16$, whereas we only show the results for $m = 6, 8, 10$ in Fig.~\ref{fig4}.  It is noted that the stability properties of the PHEOD siltions and their stability eigenvalue spectrum, are all  similar to the PQS as described in Ref.~\cite{Tam2019OL}. 

\section{CONCLUSION}
In summary, we apply the variational approach to investigate the properties of PHEOD solitons. The general expression of Lagrangian for different dispersion order are given, and the approximate analytical solutions for PHEOD solitons are obtained. The key PHEOD soliton parameters are compared between analytical and numerical results. Is is shown at low-order dispersion ($m<8$), our analytical results are essentially consistent with the numerical results. The linear stability analysis shows that all the PHEOD solitons  can be transmitted stably and have internal modes associated with soliton propagation. The results we obtain deepen the physical understanding of the evolutionary properties of PHEOD solitons.

\section*{Acknowledgements}
This work was supported by the Guangdong Basic and Applied Basic Research Foundation (Grant Nos. 2023A1515012432).

\end{document}